\documentclass[10pt,twocolumn,twoside]{IEEEtran}
\pdfoutput=1
\IEEEoverridecommandlockouts

\usepackage{cite}
\usepackage{amsmath,amssymb,amsfonts,amsthm, bm}
\usepackage{algorithm,algpseudocode}
\usepackage{graphicx}
\usepackage{paralist}
\usepackage{multirow}

\newtheorem{theorem}{Theorem}

\usepackage[breaklinks, colorlinks, citecolor=blue]{hyperref}

\newcommand{\bbS}{\mathbb{S}}

\def\bbR{\mathbb R}

\def\bbS{\mathbb S}

\def\bbE{\mathbb E}

\newcommand{\cA}{\mathcal{A}}
\newcommand{\cB}{\mathcal{B}}
\newcommand{\cC}{\mathcal{C}}

\newcommand{\cE}{\mathcal{E}}

\newcommand{\cG}{\mathcal{G}}
\newcommand{\cH}{\mathcal{H}}

\newcommand{\cJ}{\mathcal{J}}
\newcommand{\cK}{\mathcal{K}}
\newcommand{\cL}{\mathcal{L}}

\newcommand{\cN}{\mathcal{N}}
\newcommand{\cO}{\mathcal{O}}

\newcommand{\cS}{\mathcal{S}}
\newcommand{\cT}{\mathcal{T}}
\newcommand{\cU}{\mathcal{U}}
\newcommand{\cV}{\mathcal{V}}

\newcommand{\cX}{\mathcal{X}}
\newcommand{\cY}{\mathcal{Y}}

\newcommand{\bE}{\mathbf{E}}

\newcommand{\bJ}{\mathbf{J}}

\newcommand{\bU}{\mathbf{U}}

\newcommand{\Q}{\mathds{Q}}

\DeclareMathOperator*{\argmax}{argmax}
\DeclareMathOperator*{\argmin}{argmin}
\DeclareMathOperator*{\argopt}{argopt}

\newcommand{\true}{\text{\textbf{True}}}
\newcommand{\false}{\text{\textbf{False}}}

\newcommand{\BREAK}{\textbf{Break}}

\newcommand{\mbfR}{\mathbf{R}}

\newcommand{\mbfI}{\mathbf{I}}

\newcommand{\mbfone}{\mathbf{1}}
\newcommand{\mbfzero}{\mathbf{0}}

\newcommand{\trace}[1]{\text{tr}({#1})}
\newcommand{\diag}[1]{\text{diag}({#1})}

\newcommand{\SWITCH}[1]{\STATE \textbf{switch} {#1} \textbf{:} \begin{ALC@g}}
\newcommand{\ENDSWITCH}{\end{ALC@g}}
\newcommand{\CASE}[1]{\STATE \textbf{case} {#1} \textbf{:} \begin{ALC@g}}
\newcommand{\ENDCASE}{\end{ALC@g}}

\newcommand{\mat}[1]{\begin{bmatrix} #1 \end{bmatrix}}

\newcommand{\card}[1]{{\lvert #1 \rvert}}
\newcommand{\vecsqnorm}[2]{\lVert {#1} \rVert_{#2}^2}

\newcommand{\uniformcontinuous}[1]{}

\newcommand{\bcU}{\bm{\cU}}

\def\BibTeX{{\rm B\kern-.05em{\sc i\kern-.025em b}\kern-.08emT\kern-.1667em\lower.7ex\hbox{E}\kern-.125emX}}

\begin{document}

\title{Self-Tuning Network Control Architectures with Joint Sensor and Actuator Selection}

\author{Karthik Ganapathy, Iman Shames, Mathias Hudoba de Badyn, and Tyler Summers
\thanks{This work was sponsored by the United States Air Force Office of Scientific Research under award number FA2386-19-1-4073 and by the National Science Foundation under award number ECCS-2047040.}
\thanks{K. Ganapathy and T. Summers are with the Control, Optimization and Networks Lab, The University of Texas at Dallas (email: \{\href{mailto:karthik.ganapathy@utdallas.edu}{karthik.ganapathy}, \href{mailto:tyler.summers@utdallas.edu}{tyler.summers}\}@utdallas.edu)}
\thanks{I. Shames is with the CIICADA Lab, Australian National University, Canberra. (email: \href{mailto:iman.shames@anu.edu.au}{iman.shames@anu.edu.au})}
\thanks{M. Hudoba de Badyn is with the Department of Technology Systems (ITS) at the University of Oslo, Norway. (email: \href{mailto:mathias.hudoba@its.uio.no}{mathias.hudoba@its.uio.no})}
}

\maketitle
\thispagestyle{empty}

\begin{abstract}
We formulate a mathematical framework for designing a self-tuning network control architecture, and propose a computationally-feasible greedy algorithm for online  architecture optimization. In this setting, the locations of active sensors and actuators in the network, as well as the feedback control policy are jointly adapted using all available information about the network states and dynamics to optimize a performance criterion. We show that the case with full-state feedback can be solved with dynamic programming, and in the linear-quadratic setting, the optimal cost functions and policies are piecewise quadratic and piecewise linear, respectively. Our framework is extended for joint sensor and actuator selection for dynamic output feedback control with both control performance and architecture costs. For large networks where exhaustive architecture search is prohibitive, we describe a greedy heuristic for actuator selection and propose a greedy swapping algorithm for joint sensor and actuator selection. Via numerical experiments, we demonstrate a dramatic performance improvement of greedy self-tuning architectures over fixed architectures. Our general formulation provides an extremely rich and challenging problem space with opportunities to apply a wide variety of approximation methods from stochastic control, system identification, reinforcement learning, and static architecture design for practical model-based control.
\end{abstract}

\section{Introduction}

Emerging complex dynamical networks present tremendous challenges and opportunities to fundamentally re-imagine their control architectures and control algorithms \cite{annaswamy2023ControlSocietalscaleChallenges}. In large-scale networks, the structure of the control architecture (here defined as the locations of sensors, actuators, and their communication patterns) is crucial to their performance and robustness properties and an important design consideration. Much of control theory operates with fixed control architectures, with the design focused almost entirely on the control policy rather than the architecture. There is a long history of adaptive control and reinforcement learning of adapting policy parameters online to measured data and/or identified models, but these ideas have never been applied to the architecture itself in a feedback loop with measured data. Here we propose \textit{self-tuning network control architectures} that jointly adapt the policy \textit{ and } the architecture online to measured data, analogous to the classical self-tuning regulator in adaptive control \cite{astrom2013AdaptiveControlSecond}.

Such self-tuning architecture is compelling for large infrastructure networks and complex, high-dimensional networks with time-varying phenomena, such as power grids with massive penetration of inverter-based resources, mixed-mode transportation networks, epidemic/information spread in social networks, and economic activity in large economies. Integration of new sensing and actuation technologies offers an increasingly large number of points from which to estimate and control complex dynamic phenomena. However, this influence is restricted by resource and budget constraints that affect the number and performance of sensors and actuators at a given time. Self-tuning architecture can provide improved performance and robustness over fixed architecture for networks and dynamic systems under such limitations.

The vast growing literature on adaptive control \cite{astrom2013AdaptiveControlSecond} and reinforcement learning (RL) \cite{sutton2018ReinforcementLearningSecond, bertsekas2019ReinforcementLearningOptimal} focuses on adapting policy parameters based on measured data. The self-tuning regulator \cite{astrom2013AdaptiveControlSecond} is a prototypical adaptive control approach in which model parameters related to system dynamics are estimated online from data and then used to adapt the parameters of a feedback policy through a control design procedure. This approach can flexibly accommodate many combinations of parameter estimation and control design methodologies. There are several other approaches to adaptive control and RL of indirect and model-free flavors that adapt policy parameters through the estimation of other quantities, such as value functions and policy gradients. However, most similar work focus on individual systems, and work in a network context utilizes architecture with fixed locations for sensors and actuators.

Approaches to the design of control architectures by optimizing controllability and observability metrics have received considerable attention in recent years \cite{olshevsky2014MinimalControllabilityProblems, pasqualetti2014ControllabilityMetricsLimitations, tzoumas2016MinimalActuatorPlacement, summers2016SubmodularityControllabilityComplex, ruths2014ControlProfilesComplex, hudobadebadyn2021H2PerformanceSeriesparallel, foight2020PerformanceDesignConsensus, ganapathy2022ActuatorSelectionDynamical}. However, architecture design is largely treated as a single static design problem. Some works have studied time-varying models of actuator scheduling \cite{zhao2016SchedulingControlNodes, nozari2017TimeinvariantTimevaryingActuator,
siami2021DeterministicRandomizedActuator, olshevsky2020RelaxationTimeVaryingActuator, siami2020SeparationTheoremJoint}, but the architecture, while time-varying, remains open loop and does not adapt to changing network state or dynamics. Very recent work has considered selection actuators for uncertain systems based on data measured from the system in limited settings with linear systems and specific controllability metrics \cite{fotiadis2021LearningBasedActuatorPlacement, ye2023OnlineActuatorSelection}.

Our main contributions are summarized as follows.
\begin{enumerate}
    \item We formulate a general mathematical framework for self-tuning network control architecture design and propose a general solution structure analogous to the classical self-tuning regulator from adaptive control. (Section \ref{sec:architecture_optimization_problem_general_framework})
    \item For the special case of full-state feedback and known dynamics, we show that dynamic programming can solve the problem in principle, which couples feedback policy design with a search over the combinations of architecture. In the linear quadratic setting, we show that optimal cost functions and policies are piecewise quadratic and affine, respectively. (Section \ref{sec:architecture_optimization_problem_dp_fullfeedback})
    \item We propose a computationally tractable greedy heuristic for self-tuning LQG architectures. The greedy algorithm is further modified to consider the sequential selection and rejection of the architecture for computational efficiency. (Section \ref{sec:algorithm})
    \item In numerical experiments, we demonstrate the efficacy of our proposed framework and selection algorithms and show how self-tuning architecture can provide improved performance over a fixed architecture. (Section \ref{sec:empirical_analysis})
\end{enumerate}

A preliminary version of this work appeared in \cite{summers2022SelfTuningNetworkControl}, which focused only on the actuator optimization problem under full-state feedback and described a basic greedy heuristic. Here, we significantly extend this preliminary version in three ways. First, we formulate joint sensor and actuator selection problems with output feedback that accounts for costs related to both state estimation and feedback control. Second, in this context we develop more sophisticated greedy algorithms that feature sequential selection and rejection to improve upon simple greedy heuristics. Third, we present additional extensive numerical experiments that demonstrate dramatic performance improvements over fixed architectures.



\subsection{Notation}

The real-valued vectors and matrices in the $n$-dimensional space are denoted by $\bbR^n$ and $\bbR^{n\times n}$, respectively. The element in the $i^\text{th}$ row and $j^\text{th}$ column of matrix $A$ is denoted by $A = [a_{i,j}]$. The canonical basis vectors of $\bbR^n$ are denoted by $\mathbf{e}_i\>\forall i\in[1,n]$ where $\mathbf{e}_i$ is a vector of zeros where the $i^{th}$ element is 1. The set of symmetric positive definite matrices with real values is denoted by $\bbS^n_{++}$ where $A\in\bbS^n_{++}\Leftrightarrow A\succ0$. The set of positive semidefinite symmetric matrices is denoted by $\bbS^n_+$ where $A\in\bbS^n_+\Leftrightarrow A\succeq0$. The $n$ dimensional identity matrix is denoted by $\mbfI_n$. The vector of ones in $\bbR^n$ is denoted by $\mbfone_{n}$, the vector of zeros in $\bbR^n$ is denoted by $\mbfzero_{n}$, and the matrix of zeros in $\bbR^{n\times m}$ is denoted by $\mbfzero_{n\times m}$. The expectation of a random variable $x$ is denoted by $\bE[x]$. The Kronecker delta $\delta_{i,j} = 1$ if $i=j$, and $0$ otherwise. The uniform continuous distribution over an interval $(a,b)$ is denoted by $\bcU_{(a,b)}$. We define $\vecsqnorm{x}{W} = x^\top W x$ for a vector $x \in \bbR^n$ with a positive semidefinite weight matrix $W \in \bbS^n_+$. If the norm is unweighted, that is, if $W=\mbfI_n$ of a suitable dimension, then the subscript is dropped. For two subsets $S_1, S_2 \subseteq \cS$, let $\card{S_1}$ denote the cardinality of $S_1$, and $S_1^c$ denote the complement of the set. Let $\argopt_{x\in\cX}f(x)$ be the optimal argument $x\in\cX$ of a function $f(x)$, to represent the operators $\argmin$ or $\argmax$ for minimization or maximization, respectively.


\section{Architecture Optimization Problem} \label{sec:architecture_optimization_problem}

\subsection{General Framework} \label{sec:architecture_optimization_problem_general_framework}

We first formulate a general mathematical framework for self-tuning network control architectures. Consider a dynamical network with an underlying graph $\cG=(\cV,\cE(t))$, where $\cV=\{1,\dots,n\}$ is a set of nodes and $\cE(t)\subseteq\cV\times\cV$ is a set of time-varying edges connecting these nodes over a discrete time horizon $t\in[0,\dots,\cT]$. We associate a variable $x_{i}(t)\in\cX_i$ with node $i\in\cV$. The network state is $x(t)=[x_{1}(t), x_{2}(t), \dots, x_{n}(t)]\in\cX=\prod_{i=1}^n\cX_i$. The edges represent dynamical interactions between the nodal states.

We define a finite set of possible actuator locations in the network as $\cB=\{u_1, u_2, \dots, u_m\}$, each of which corresponds to an input signal $u_{i}(t)\in\cU_i$ that affects the state dynamics. Similarly, we define a finite set of possible sensor locations $\cC=\{y_1, y_2, \dots, y_l\}$, each of which corresponds to a distinct output measurement $y_{i}(t)\in\cY_i$ related to the state of the network. For a subset of actuators $\cA_t\subseteq\cB$ and sensors $S_t\subseteq\cC$ that are active (in operation) at time $t$ and input and output signals $u^{\cA_t}(t)\in\prod_{u_i\in \cA_t}\cU_i$ and $y^{S_t}(t)\in\prod_{y_i\in S_t}\cY_i$ respectively, the network dynamics are given by
\begin{subequations}
\begin{align}
    x(t+1) &= f_{\theta_t}^{\cA_t}\left(x(t), u^{\cA_t}(t), w(t)\right)\\
    y^{S_t}(t) &= h_{\theta_t}^{S_t}\left(x(t), v(t)\right),
\end{align}\label{eq:general_dynamics}
\end{subequations}
where $w(t)$ is an i.i.d.~stochastic process disturbance from a distribution function $P_w^{\theta_t}$, $v(t)$ is an i.i.d~measurement noise from a distribution $P_v^{\theta_t}$, and $\theta_t$ is a (generally) unknown and time-varying dynamics parameter specifying the dynamical map $f_{\theta_t}^{\cA_t}$ and measurement map $h_{\theta_t}^{S_t}$. In general, the goal is to jointly adapt the active sensors, actuators, and control inputs to all available information about the state and dynamics parameters of the network over time to optimize a performance objective. Certain network states and dynamics may render a fixed network control architecture ineffective, which motivates adapting not just the input signal, but also the architecture. When the dynamics parameter is unknown, this may involve a combination of statistical estimation/identification of $\theta_t$, control design based on an estimate of $\theta_t$, and a representation of the estimation uncertainty.

The system in \eqref{eq:general_dynamics} describes a network with changing dimensions of control input and measurement space dimensions, which are subsets of the spaces of all possible control inputs and measurement respectively, i.e., $\prod_{u_i\in \cA_t}\cU_i \subseteq \prod_{u_i\in \cB_t}\cU_i$ and $\prod_{y_i\in S_t}\cY_i\subseteq\prod_{y_i\in \cC_t}\cY_i$. This can also be represented as a system with fixed dimensions for inputs and measurements, with components that can be turned on or off. Let the active architecture sets $\cA_t, S_t$ at every time step be indicated by $1$s at the corresponding $i^{th}$ indices of the binary indicator vectors $\cA'_t \in \{0,1\}^\card{\cB}$ for actuators and $S'_t \in \{0,1\}^\card{\cC}$ for sensors, respectively. Then $u^{\cA_t}(t) = \diag{\cA'_t}u^{\cB}(t)$ and $y^{S_t}(t) = \diag{S'_t}y^{\cC}(t)$ to set the inactive actuators and sensors to $0$. 






\paragraph{The cardinality-constrained architecture design problem}
\sloppy Let $\cB_\cK = \{\cA\in2^\cB \mid \card{\cA} = \cK\}$ and $\cC_\cL =  \{S \in 2^\cC \mid \card{S} = \cL\}$ denote all possible subsets of actuators and sensors with cardinality $\cK$ and $\cL$, respectively. Let $y_{0:t} = \left[ y^{S_0}(0), y^{S_1}(1), \dots, y^{S_t}(t) \right] \in\cY_{0:t}$ and $u_{0:t-1} = \left[ u^{\cA_0}(0), u^{\cA_1}(1), \dots, u^{\cA_{t-1}}(t-1) \right] \in \cU_{0:t-1}$ denote the output and input histories, respectively. We define an \textit{architecture policy} $\pi_t : \cY_{0:t} \times \cU_{0:t-1} \rightarrow \cC_\cL \times \cB_\cK \times \prod_{u_i \in \cA_t}\cU_i$ with $\cA_t \subseteq \cB_\cK$ as a mapping from the input-output history to the sensor and actuator subsets and the next input and output so that $u^{\cA_t}(t)=\overline{\pi}_t(y_{0:t}, u_{0:t-1})$ and $y^{S_{t+1}}(t+1) = h_{\theta_{t+1}}^{S_{t+1}} \left(x(t+1), v(t+1)\right)$. Thus, an architecture policy specifies
\begin{inparaenum}[(a)]
\item a set of active sensors and actuators at each time, and
\item a feedback policy specifying the control input for active actuators at each time.
\end{inparaenum}
Mathematically, $\pi_t$ defines the triple $(\cA_t, S_{t+1}, \overline{\pi}_t)$ which specifies which actuators to use at time $t$, which measurements to collect at time $t$, and which input values to apply at time $t$. Each component of $(\cA_t, S_{t+1}, \overline{\pi}_t)$ depends on the available information (i.e., the dynamics, measurements, and available actuator and sensor locations). For a finite time horizon of length $\cT$, we define $\pi=[\pi_0, \pi_1, \dots, \pi_{\cT-1}]$.

For a given architecture policy $\pi$, with $u^{\cA_t}(t)=\overline{\pi}_t(y_{0:t}, u_{0:t-1})$, we define a cost function for initial state $x(0)=x$ as
\begin{align*}
    J_{\pi}(x) = \bE_{w,v} \left[ \sum_{t=0}^{\cT-1} c_t(x(t), u^{\cA_t}(t)) + c_{\cT}(x(\cT)) \right],
\end{align*}
where $c_t:\cX \times \prod_{u_i \in \cA_t}\cU_i \rightarrow \bbR$ is a stage cost function, $c_\cT: \cX \rightarrow \bbR$ is a terminal cost function and expectation is taken with respect to the process disturbance and measurement noise sequences. The self-tuning network architecture design problem is then to find an optimal architecture-policy, i.e.
\begin{align}
    J^*(x) = \min_{\pi}J_\pi(x), \quad \pi^*\in \argmin_\pi J_\pi(x). \label{eq:optimal_arch_policy}
\end{align}

This problem poses several challenges, as it combines already-challenging (i.e., stochastic, nonlinear, output-feedback, data-driven) feedback control design with a combinatorial architecture search. Generally, this will require approximation and heuristics for both control and architecture design. Nevertheless, we believe that this general formulation provides an extremely rich problem space with many exciting possibilities for applying a wide variety of tools from stochastic control, system identification, reinforcement learning, and static network control architecture design.

\begin{figure}[ht!]
  \centering
  \includegraphics[width=0.8\columnwidth]{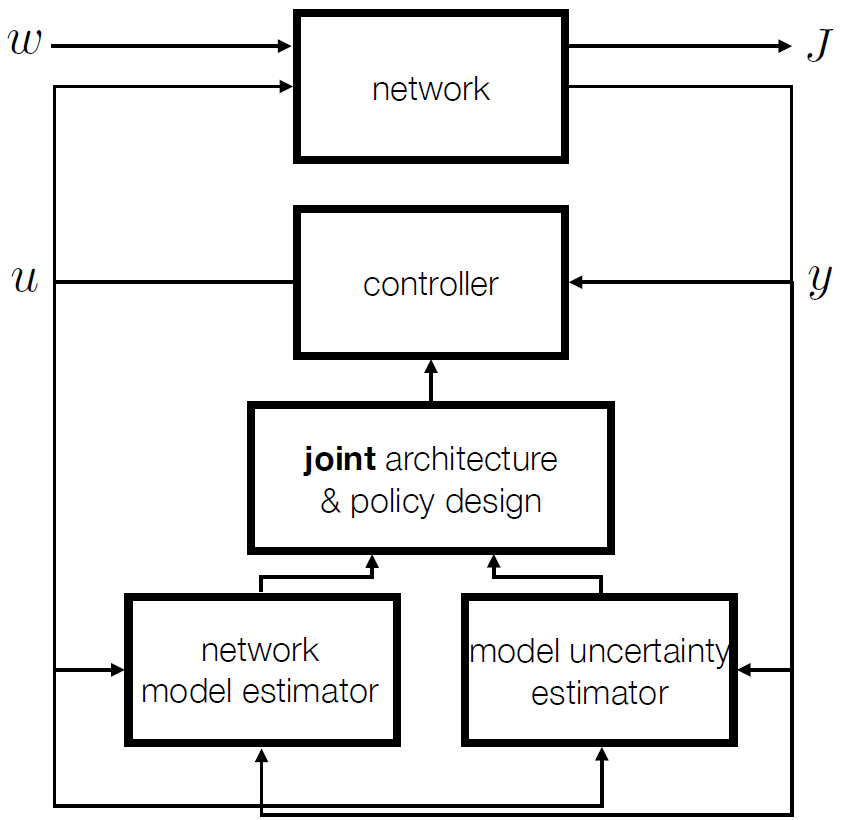}
  \caption{A self-tuning network control architecture flowchart}
  \label{fig:selftuning_flowchart}
\end{figure}

\paragraph{Self-tuning architectures}
A block-diagram illustrating the flow of a general self-tuning network architecture is shown in Fig. \ref{fig:selftuning_flowchart}, which is analogous to a self-tuning regulator from adaptive control. The architecture can be viewed as a pair of coupled loops. An inner loop features the dynamical network in closed loop with a feedback controller. An outer loop estimates the model parameters $\theta_t$ and a model uncertainty representation (such as a probability distribution or an uncertainty set) based on input-output data, which are then fed into a \textit{joint} architecture-controller design module that computes the optimal sensor and actuator subsets together with a feedback policy. Our formulation adapts the architecture and policy parameters at the same rate, but it is possible (and potentially computationally advantageous) to use different rates for each of the model estimation, policy parameter adaptation, and architecture adaptation. For example, architecture adaptation may occur at a much slower rate than model estimation or policy parameter updates while still providing substantial performance benefits. This approach can also be viewed as maintaining a data-driven digital twin of the network and adapting both the architecture and policy as the digital twin evolves \cite{tao2019DigitalTwinIndustry}.

The proposed self-tuning architecture problem is highly flexible in terms of the underlying methods for system identification, feedback control design, and network control architecture design. While many such strategies have been explored in the literature in the context of fixed architecture design, to the best of our knowledge no prior work has considered the online adaptation of the control/sensing architecture in conjunction with the control input design.

\subsection{Dynamic Programming for Full State Feedback} \label{sec:architecture_optimization_problem_dp_fullfeedback}

We consider a special case where the full network state measurement is available for feedback, and the architecture design problem consists of selecting only actuator subsets at each time based on the current state and dynamics parameters and their estimates. Let $x_{0:t} = [x(0), x(1), \dots, x(t)] \in \cX_{0:t}$ be the history of the network state. An architecture policy $\pi_t$ defines an active actuator set and a feedback policy $(\cA_t, \overline{\pi}_t)$ such that $u^{\cA_t}(t) = \overline{\pi}_t(x_{0:t}, u_{0:t-1})$. In general, the self-tuning architecture involves using the available state input data $(x_{0:t}, u_{0:t-1})$ to estimate the dynamics parameter $\theta_t$, and then selecting an architecture policy through a control design method.

When the dynamics parameter $\theta_t$ is known, the Markovian structure allows consideration of architecture-policies of the current state $u^{\cA_t}(t) = \overline{\pi}(x(t))$. This then permits a dynamic programming solution to the problem stated below.

\begin{theorem}
Consider the optimal control problem
\begin{align}
    \min_{\pi} \> &\bE_{x_0,w} \left[\sum_{t=0}^{\cT-1} c_t \left( x(t), u^{\cA_t}(t) \right) + c_{\cT}(x(\cT))\right] \label{eq:thm1_problem} \\
    \text{subject to} \>\>\> &x(t+1) = f^{\cA_t}_{\theta_t} \left( x(t), u^{\cA_t}(t), w(t) \right). \nonumber
\end{align}
The optimal cost function $J^*(x)$ \eqref{eq:optimal_arch_policy} is obtained from the last step of the dynamic programming recursion
\begin{align}
    &J_\cT(x) = c_\cT(x) \label{eq:thm1_terminal}\\
    &J_t(x) = \min_{\genfrac{}{}{0pt}{}{(\cA_t, u)\in \cB_\cK \times}{\prod_{u_i \in \cA_t}\cU_i}} \bE_w \left[ c_t(x,u^{\cA_t}) + J_{t+1}(f_{\theta_t, \cA_t}(x, u^{\cA_t}, w)) \right] \label{eq:thm1_opt_cost}
\end{align}
and corresponding optimal architecture policies are obtained via
\begin{align}
    \pi^*_t(x) \in \argmin_{\genfrac {}{}{0pt}{}{(\cA_t, u)\in \cB_\cK \times}{\prod_{u_i \in \cA_t}\cU_i}} \bE_w \left[ c_t(x,u^{\cA_t}) + J_{t+1}(f_{\theta_t, \cA_t}(x, u^{\cA_t}, w)) \right].
    \label{eq:thm1_opt_policy}
\end{align}
\end{theorem}
\begin{proof}
    The proof is provided in Appendix \ref{proof:thm1}.
    
\end{proof}

\subsubsection{Self-tuning LQR Architectures}
We now consider a special case with linear dynamics and quadratic cost functions. Nodal states are vectors $x_{i,t}\in\cX_i = \bbR^{n_i}$, and the network state is then $x(t) \in \cX = \bbR^N$ with $N = \sum_i n_i$. The inputs are also vectors $u_{i}(t) \in \cU_i = \bbR^{m_i}$. In this setting, the set of possible actuator locations can be identified with a set of columns that can be used to construct an input matrix. In particular, we can set $\cB = \{b_1, b_2, \dots, b_M\}$, where $b_i \in \bbR^N$. For $\cA_t \in \cB$, we form the input matrix
\begin{align*}
    B^{\cA_t} = \mat{
        | & | & \dots & | \\
        b_{i_1} & b_{i_2} & \dots & b_{i_k} \\
        | & | & \dots & | \\
    } \in \bbR^{N \times |\cA_t|}.
\end{align*}
For $|\cA_t|=\cK$, there are $\binom{n}{\cK}$ possible input matrices, up to permutation of the columns. We can identify this set with $\cB_\cK$ so that $\cB_\cK = \{B^{\cA_t} \in \bbR^{N\times|\cA_t|} \mid \cA_t \subset\cB, \card{\cA_t}=\cK\}$. The system dynamics with active input matrix $B^{\cA_t}$ at time $t \in \{0, \dots, \cT-1\}$ is given by
\begin{align}
    x(t+1) = Ax(t) + B^{\cA_t} u^{\cA_t}(t) + w(t),
\end{align}
where $w(t)$ is an i.i.d.~$\bbR^n$-valued zero-mean random vector with covariance matrix $W$. The following result establishes a more explicit structure for optimal cost functions and architecture policies.

\begin{theorem}
Consider the linear quadratic optimal control problem over state feedback architecture policies $\pi$ given by
\begin{align}
    \min_\pi \> &\bE_{x_0,w} \left[\sum_{t=0}^{\cT-1} \vecsqnorm{x(t)}{Q_t} + \vecsqnorm{u_t}{R_t} \right] + \vecsqnorm{x(\cT)}{Q_\cT} \nonumber\\
    \text{subject to} \>\>\> & x(t+1) = Ax(t) + B^{\cA_t} u^{\cA_t}(t) + w(t) \nonumber\\
    &B^{\cA_t} \in \cB_\cK. \label{eq:thm2_problem}
\end{align}
The optimal cost-to-go functions and policies obtained from the dynamic programming recursion \eqref{eq:thm1_terminal}, \eqref{eq:thm1_opt_cost} and \eqref{eq:thm1_opt_policy} for the problem \eqref{eq:thm2_problem} are piecewise quadratic and piecewise linear respectively, defined over a finite partition of the state-space $\bbR^n$.
\end{theorem}
\begin{proof}
The proof is provided in Appendix \ref{proof:thm2}.
\end{proof}
This result provides an exact algorithm to construct the optimal cost-to-go functions and policies. A simple numerical example with a 2-dimensional state space illustrating an optimal architecture policy is provided in our preliminary work \cite{summers2022SelfTuningNetworkControl}. However, an exhaustive search over all available architectures is computationally intractable for all but the actuator sets of the smallest cardinality. This motivates approximation algorithms for the optimization of joint architecture policies. This is discussed in the next section.

\subsection{Self-tuning LQG Architectures with Estimation and Control for Output Feedback Optimal Control} \label{sec:architecture_optimization_problem_dp_partialfeedback}

We extend the optimization problem to the sequential selection of both actuators and sensors for optimal control. This problem is challenging because the separation principle fails for joint architecture-policy design. For fixed architectures, the policy design problem separates into a backward-in-time recursion for computing a feedback control law and a forward-in-time recursion for propagating state estimates. However, for self-tuning architectures, both the controller and the state estimator depend on which sensors and actuators are selected. Future actuator and control decisions require accurate estimates of the past sensor decisions and the estimated state trajectory. We address this by assessing costs over a finite length, receding time horizon where the architecture is fixed, analogous to a Model Predictive Control (MPC) approach. The framework predicts cost over the horizon as a function of the current state and estimation error covariance to predict the effect of different architectures. Over a finite time horizon, the costs are finite and provide a scalar metric of comparison for selecting architectures.


\subsubsection{Network dynamics}
The set of active actuator indices at time step $t$ is given by $m_t = \{ i \mid \cA'_{t,i}=1, i\in[1,m] \}$ and the number of active actuators is given by $|m_t|={\cA'}_t^\top {\cA'}_t$ at a time step $t$. Similarly, we can define $l_t = \{ i \mid S'_{t,i}=1, i\in[1,l] \}$ and $|l_t|={S'}_t^\top {S'}_t$ for the set of indices and the number of active sensors in a particular time step. The corresponding actuator matrix $B^{\cA_t}$ and sensor matrix $C^{S_t}$ are formed by the columns of the active architecture in $\cB$ indexed by $m_t$ and the rows in $\cC$ indexed by $l_t$. The dynamics of the system are given by
\begin{align*}
    x(t+1) &= A x(t) + B^{\cA_t} u^{\cA_t}_t + w(t)\\
    y^{S_t}(t) &= C^{S_t} x(t) + v(t)
\end{align*}
where $y^{S_t}(t)$ is the set of measurements and $u^{\cA_t}_t$ is the set of control inputs at that time step $t$. Note that the control inputs and measurements are functions of the active actuator and sensor sets at the current time step as in \eqref{eq:general_dynamics}, and we are dropping the subscripts $\cA_t, S_t$ for simplicity of notation. The additive process noise $w(t)$ and measurement noise $v(t)$ are i.i.d. zero mean with $\bE[w(t) w(t)^\top] = W \in \bbS_+^n$ and $\bE[v(t) v(t)^\top] = V^{S_t} \in \bbS_+^\card{S_t}$, respectively. The measurement noise covariance is a function of $S_t$ for both values and dimensions. However, we assume that each sensor noise is independent with identical variance and drop the subscript for notation convenience.

For a given sensor subset, a state estimator has the form
\begin{subequations}
\begin{align*}
    \hat{x}(t+1) &= A(\mbfI - L^{S_t} C^{S_t})\hat{x}(t) + A L^{S_t} y^{S_t}(t) + B^{\cA_t} u^{\cA_t}(t)\\
    u^{\cA_t}(t) &= -K^{\cA_t} \hat{x}(t)
\end{align*}
\end{subequations}
where $\hat{x}(t)$ is the state estimate, $L^{S_t}$ is an estimation gain and $K^{\cA_t}$ is a feedback gain at time step $t$ corresponding to the active actuators and sensors, which are described next.

\subsubsection{Finite horizon output feedback control}


For fixed architectures, it is well known that the optimal linear quadratic Gaussian (LQG) controller consists of a Kalman filter composed with a linear quadratic regulator, which we briefly summarize here.

\paragraph{Control}
The optimal feedback control problem aims to minimize the expected cost
\begin{align*}
    J_u = \min_{u(.)} \bE \left[ \sum_{\tau=0}^{\cT-1} \left( \vecsqnorm{x(\tau)}{Q_{\cB}} + \vecsqnorm{u^{\cA_t}(\tau)}{R_{1,\cB}} \right) + \vecsqnorm{x(\cT)}{Q} \right]
\end{align*}
where $Q\in\bbS_{+}^n$ is the cost on states and $R_{1,\cB}\in\bbS_{+}^{|m_t|}$ is the cost on active actuators. For a fixed architecture, the optimal feedback policy is given by $u^{\cA_t}(\tau) = -K_{\tau}^{\cA_t}\hat{x}(\tau)$ where the gains $K_{\tau}^{\cA_t}$ and cost matrices $P_\tau^{\cA_t}$ are computed from the backwards recursion
\begin{subequations}
\begin{align*}
    K_{\tau}^{\cA_t} &=  \left( {B^{\cA_t}}^\top P_{\tau+1}^{\cA_t} B^{\cA_t} + R_{1, \cB} \right)^{-1} {B^{\cA_t}}^\top P_{\tau+1}^{\cA_t} A\\
    P_{\tau}^{\cA_t} &= A^\top P_{\tau+1}^{\cA_t} A - A^\top P_{\tau+1}^{\cA_t} B^{\cA_t} ({B^{\cA_t}}^\top P_{\tau+1}^{\cA_t} B^{\cA_t} \nonumber \\ & \quad\quad+ R_{1, \cB} )^{-1} {B^{\cA_t}}^\top P_{\tau+1}^{\cA_t} A + Q_{\cB}
\end{align*}
\end{subequations}
over the prediction horizon $\tau\in[0,\cT-1]$, initialized with $P_{\cT}^{\cA_t} = Q$, and $\hat{x}(\tau)$ is the state estimate obtained from the optimal state estimator, the Kalman filter.

\paragraph{State Estimation}
The estimator filters the output measurements to form an estimate of the state. The optimal estimation gains $L_{\tau}^{S_t}$ and the error-covariance matrices $E_{\tau}^{S_t}$ over the prediction horizon $\tau\in[0,\cT-1]$ are computed by the forwards recursion
\begin{subequations}
\begin{align*}
    L_{\tau}^{\cA_t} &= E_{\tau}^{S_t} {C^{S_t}}^\top \left( C^{S_t} E_{\tau}^{S_t} {C^{S_t}}^\top + R_{1, \cC}^{S_t} \right)^{-1}\\
    E_{\tau+1}^{S_t} &= A E_{\tau}^{S_t} A^\top - A E_{\tau}^{S_t} {C^{S_t}}^\top (C^{S_t} E_{\tau}^{S_t} {C^{S_t}}^\top \nonumber \\ & \quad\quad+ R_{1, \cC})^{-1} C^{S_t} E_{\tau}^{S_t} A^\top + W.
\end{align*}
\end{subequations}
where $E_{\tau=0}^{S_t} = E_{t}^{S_t}$ is the estimation error covariance at the current time step $t$ and $R_{1, \cC}^{S_t} = V^{S_t}$ is the covariance of measurement noise in the active sensors.

\paragraph{Closed-loop Augmented System}
The closed-loop system dynamics with optimal control and estimation gains is
\begin{subequations}
\begin{align*}
    \mat{x(t+1) \\ \hat{x}(t+1)} &= \mat{ A & -B^{\cA_t} K^{\cA_t}_0 \\ AL_{0}^{S_t}C^{S_t} & A - AL_{0}^{S_t}C^{S_t} - B^{\cA_t}K_{0}^{\cA_t}} \mat{x(t) \\ \hat{x}(t)} \nonumber \\ & \quad\quad + \mat{ \mbfI & \mbfzero \\ \mbfzero & AL_{0}^{S_t} } \mat{w(t) \\ v(t)}  
\end{align*}
\end{subequations}
Let the stacked vector of the states and its estimates be $X(t) = \mat{x^\top(t) & \hat{x}^\top(t) }^\top \in\bbR^{2n}$, the dynamics matrix of this closed-loop system be $\overline{A}_{t,\tau} \in \bbR^{2n\times 2n}$, and the noise be $W(t) = F_{t,0} \mat{w(t)^\top & v(t)^\top }^\top$.

As the true state $x(t)$ is unknown to the controller, the predicted cost $J_{t,\cT}$ for the $\cT$-step horizon from the current time $t$ is evaluated as a function of the estimate of the states $\hat{X}(t) = \mat{\hat{x}^\top(t) & \hat{x}^\top(t) }^\top$. It is defined as $\cJ^{\text{est}}$ by
\begin{align}
    \cJ^{\text{est}}_{t,\cT}(\hat{X}(t), (\cA_t, S_t)) = \vecsqnorm{\hat{X}(t)}{Z_0} + \sum_{\tau=0}^{\cT-1} tr\left( Z_{\tau+1} \overline{W}_\tau \right). \label{eq:estimated_control_cost}
\end{align}
The cost matrix $Z_\tau$ on the augmented matrix is iteratively calculated by the backwards recursion over the prediction horizon $\tau\in[0,\cT-1]$ by
\begin{align*}
    Z_{\tau} &= \overline{A}_{\tau,t}^\top Z_{\tau+1}\overline{A}_{\tau,t} + \overline{Q}_{\tau}\\
    \text{where}\>\> Z_{\cT} &= \overline{Q}_{\cT} = \mat{Q_{\cT} & \mbfzero \\ \mbfzero & \mbfzero}.
\end{align*}
The cost matrix and additive disturbance matrix on the augmented system over the prediction time horizon are defined as
\begin{align*}
    \overline{Q}_\tau &= \mat{ Q_\cB & \mbfzero \\ \mbfzero & {K_{\tau}^{\cA_t}}^\top R_{1,\cB} K_{\tau}^{\cA_t}}\\
    \overline{W}_\tau &= \bE[W(\tau) W(\tau)^\top] = F_{\tau,t} \mat{ W & \mbfzero \\ \mbfzero & V} F_{\tau,t}^\top\\
    F_{\tau,t} &= \mat{ \mbfI & \mbfzero \\ \mbfzero & AL_{\tau}^{S_t} }
\end{align*}

The true control stage cost incurred at time step $t$ is
\begin{align}
    \cJ^{\text{true}}_{t}(X(t), (\cA_t, S_t)) = \vecsqnorm{X(t)}{\overline{Q}_0} \label{eq:true_control_cost}
\end{align}

\paragraph{Architecture costs}

We allow architecture design decisions to include heterogeneous running and switching costs on the available sensors and actuators of the architecture at each time step. We consider constant running and switching costs, though it is straightforward to allow them to be time-varying.

\textbf{Running costs} are the overhead costs of running or operating active actuators and sensors. Each actuator and sensor may have specific positive costs associated with them. For available actuators $\cB$ and sensors $\cC$, we can define cost vectors $R_{2,\mathcal{B}} \in \bbR^{\card{\cB}}_{\geq 0}, R_{2,\mathcal{C}} \in \bbR^{\card{\cC}}_{\geq 0}$, respectively. Then running costs are expressed in terms of the indicator vectors $\cA'$ and $S'$ as

\begin{align} \label{eq:arch_run_cost}
    \cJ^{\text{run}}_{t} &= {\cA'}_t^\top R_{2,\mathcal{B}} + {S'}^\top_t R_{2,\mathcal{C}}.
\end{align}
\textbf{Switching costs} are the overhead costs of changing the active/inactive state of an actuator or sensor at each time step. These may be expressed in terms of the indicator vectors as 
\begin{align}\label{eq:arch_switch_cost}
    \cJ^{\text{switch}}_{t} &= (\cA'_t - \cA'_{t-1})^\top R_{3,\mathcal{B}} + (S^{'}_t - S'_{t-1})^\top R_{3,\mathcal{C}}.
\end{align}

\paragraph{Total costs}
The total \textbf{estimated cost} at time $t$ is the total of the estimated control cost over the prediction horizon \eqref{eq:estimated_control_cost} and the architecture costs. The total cumulative \textbf{true cost} at a specific time step $t$ is the cumulative total of the true stage cost \eqref{eq:true_control_cost} and the architecture costs incurred so far.
\begin{align}
    \bJ^{\text{est}}_{t}(\hat{X}(t), (\cA_t, S_t)) &= \cJ^{\text{est}}_{t} + \cJ^{\text{run}}_{t} + \cJ^{\text{switch}}_{t} \label{eq:estimated_totalcost}\\
    \bJ^{\text{true}}_{t}(X(t), (\cA_t, S_t)) &= \sum_{\tau=0}^{t} \left[ \cJ^{\text{true}}_{\tau} + \cJ^{\text{run}}_{\tau} \right] + \sum_{\tau=1}^{t} \cJ^{\text{switch}}_{\tau} \label{eq:true_totalcost}
\end{align}

\subsubsection{Architecture optimization}

We define the set of policies at time step $t$ as $\Pi_t: \cY_{0:t} \times \cU_{0:t-1} \rightarrow \{\cA_t\} \times \{S_t\} \times \cU_{t}$, mapping from the history of measurements and control inputs to the architecture and control inputs at time step $t$. The optimal policy is defined as
\begin{align}
    \pi^*_t = \argmin_{\pi_t \in \Pi_t} \bJ^{\text{est}}_{t} (\hat{X}(t), (\cA_t, S_t))
\end{align}

\section{Greedy Algorithms for Architecture Design} \label{sec:algorithm}

Since network control architecture design combines feedback control design with combinatorial architecture search, approximations and heuristic approaches are required for feasible implementation. We propose and analyze a greedy algorithm and variations thereof for online optimization of sensors and actuators.

Let $C$ be the space of all available choices
, $S\subseteq C$ be the set of active choices, $S^c \triangleq C-S \subseteq C$ be the set of inactive choices (complement of $S$ in $C$), $J:C\rightarrow\mbfR$ be a cost metric to optimize (maximize or minimize) and $\cL_m, \cL_M \in\mbfR_{>=0}$ be lower and upper bounds on the cardinality of the set.

\subsection{Greedy Selection}

Greedy selection without replacement (Alg. \ref{alg:greedy_selection}) is an iterative optimization algorithm that begins with an empty set and selects the element with the best marginal improvement of the cost metric at each iteration. The algorithm requires $\card{S^c}$ metric computations per iteration for a total of $\sum_{i=0}^{\cL_M -1} \card{C}-i$.

\begin{algorithm}[!ht]
\caption{Greedy Selection}
\label{alg:greedy_selection}
\begin{algorithmic}[1]
\Require{Available choices $C$, Metric $J$, Max constraint $\cL_{M}$}
\Ensure{Greedy optimal choices $S$}
\State $S\gets\emptyset$
\While{$\cL_M > \card{S}$}
    \State $c^* \gets \argopt_{c\in S^c} J(S + \{c\})$
    \State $S \gets S + \{c^*\}$
\EndWhile
\State \Return Greedy optimal $S$
\end{algorithmic}
\end{algorithm}

To illustrate a basic greedy heuristic, we apply this to approximate an architecture policy, focusing on self-tuning LQR architectures with a system identification component for a static dynamics matrix. The network dynamics is given by,
\begin{align}
    x(t+1) = A_\theta x(t) + B^{\cA_t} u^{\cA_t}(t) + w(t), \label{eq:parametrised_dynamics}
\end{align}
where $w(t)$ is i.i.d. zero-mean with covariance $W$. Based on the state-input history $(x_{0:t}, u_{0:t-1})$, the dynamics parameter $\theta$ is identified using a least squares estimation approach, and then a greedy algorithm is used to select a subset of actuators with cardinality $K$ to optimize the infinite-horizon LQR control performance associated with the estimated network model. In particular, the approximate architecture policy is as follows.
\begin{align*}
    u^{\cA_t}_\text{greedy}(t) = K^{\cA_t}_\text{greedy} x(t),
\end{align*}
where the computation of $K^{\cA_t}_\text{greedy}$ is computed through Alg. \ref{alg:greedy_selection_architecture}. Although this greedy algorithm is suboptimal, it renders the joint architecture policy design computationally tractable and still provides dramatically improved performance compared to fixed architectures.

\begin{algorithm}[!ht]
\caption{Greedy Actuator Selection for State Feedback Control}
\label{alg:greedy_selection_architecture}
\begin{algorithmic}[1]
\Require State-Input history $(x_{0:t}, u_{0:t-1})$ at time $t$, actuator location set $\cB = \{b_1, b_2, \dots, b_M \}$, cost matrices $Q, R$, actuator set cardinality $\cK$
\Ensure{Optimal architecture $\cA_t$ and control input $u^{\cA_{t}}_\text{greedy}$}
\State Identify dynamic parameter: $ \hat{\theta}=\argmin_{\theta} \sum_{\tau=0}^{t-1} \vecsqnorm{x(\tau+1) - (A_\theta x(\tau) + B^{S_\tau} u^{S_\tau}(\tau))}{}$.
\State Initialize $\cA_t = \emptyset, B^{\cA_t}=\mat{\quad}$.
\While{$|\cA_t| < \cK$}
    \State $s^* = \argmin_{s\in\cB} x(t)^\top P^s x(t)$ where $P^s = Q + A_{\hat{\theta}}^\top P^s A_{\hat{\theta}} - A_{\hat{\theta}}^\top P^s B^s (R + {B^s}^\top P^s B^s)^{-1} {B^s}^\top P^s A_{\hat{\theta}}$ and $B^s = \mat{B^{\cA_t} & B^s}$
    \State $\cA_t \gets \cA_t + \{s^*\}$, $B^{\cA_t} = \mat{B^{\cA_t} & b_{s^*}}$
\EndWhile
\State $P^{\cA_t} = Q + A_{\hat{\theta}}^\top P^{\cA_t} A_{\hat{\theta}} - A_{\hat{\theta}}^\top P^{\cA_t} B^{\cA_t} (R + {B^{\cA_t}}^\top P^{\cA_t} B^{\cA_t})^{-1} {B^{\cA_t}}^\top P^{\cA_t} A_{\hat{\theta}}$
\State $K^{\cA_{t}}_\text{greedy} = -(R + {B^{\cA_t}}^{\top} P^{\cA_t} B^{\cA_t})^{-1} {B^{\cA_t}}^{\top} P^{\cA_t} A_{\hat{\theta}}$
\State \Return $\cA_{t}$ and $u^{\cA_{t}}_\text{greedy} = K^{\cA_{t}}_\text{greedy} x(t)$
\end{algorithmic}
\end{algorithm}

A limitation of the greedy selection algorithm in our context is that the cost function may evaluate all options with infinity for small initial architecture sets that fail to stabilize the system.  Greedy rejection is an alternative solution and is discussed next.

\subsection{Greedy Rejection}

Greedy rejection \cite{guo2021ActuatorPlacementStructural} (Alg. \ref{alg:greedy_rejection}) is an iterative optimization algorithm that begins with the full set of available choices and removes the element with the least marginal benefit at each iteration. The algorithm has $\card{S}$ choices per iteration for a total of $\sum_{i=0}^{\cL_m-1} \card{C}-i$.

\begin{algorithm}[!ht]
\caption{Greedy Rejection}
\label{alg:greedy_rejection}
\begin{algorithmic}[1]
\Require {Available choices $C$, Metric $J$, Max constraint $\cL_{M}$}
\Ensure {Greedy optimal choices $S$}
\State $S\gets C$
\While{$\cL_{M} < \card{S}$}
    \State $c^* \gets \argopt_{c\in S} J(S - \{c\})$
    \State $S \gets S-\{c^*\}$
\EndWhile
\State \Return Greedy optimal $S$
\end{algorithmic}
\end{algorithm}

Greedy rejection addresses the limitation of greedy selection discussed in the previous section for infinite-horizon problems. With an accurate system model, a full actuator set guarantees finite costs. By setting a threshold in terms of cost, we can identify a small actuator set required to achieve this. However, this is more computationally intensive, especially for large networks with sparse architecture, and can still fail under hard constraints on the active set size.

\subsection{Greedy Swapping}

We introduce a novel swapping algorithm to address the aforementioned limitations in greedy selection and rejection algorithms. Swapping is the combinatorial problem of simultaneously switching an equal number of active and inactive choices. We decompose swapping into sequential greedy selection and rejection subsequences and propose an online greedy algorithm for joint architecture optimization with loose constraints.

Rather than starting with an empty set for selection or a full set for rejection, we initialize optimization for the current time step with the active architecture from the previous time step. We then add and remove elements (i.e., sequential swapping of active and inactive sensors and actuators) based on the known system model and trajectory information encoded in the state estimate and error covariance at the current time step. This approach can be interpreted as a combinatorial gradient descent initialized from the previously active architecture. We describe the proposed greedy swapping algorithm (lines \ref{alg_step:greedy_sequential_finite_changes}-\ref{alg_step:greedy_swap_end}) in the context of architecture optimization in Alg. \ref{alg:greedy_sequential_architecture}, where a selection subsequence (lines \ref{alg_step:greedy_sequential_selection_start}-\ref{alg_step:greedy_sequential_selection_end}) and rejection subsequence (lines \ref{alg_step:greedy_sequential_rejection_start}-\ref{alg_step:greedy_sequential_rejection_end}) are sequentially iterated over.


Let the architecture constraints be $\cL(\cA_t, S_t)$ where
\begin{align}
    \cL(\cA_t, S_t) =
    \begin{cases}
        \true & \text{ if } \cL_m\leq \card{\cA_t} \leq \cL_M \\ & \text{ and } \cL_{m'}\leq \card{S_t} \leq \cL_{M'}\\
        \false & \text{otherwise}.
    \end{cases}\label{eq:arch_limits}
\end{align}
The feasible selection and rejection options depend on the cardinality of the active architecture sets, and their hard lower and upper constraints, denoted as $\cL_m, \cL_M$ for actuators and $\cL_{m'}, \cL_{M'}$ for sensors.

Let $\cO:(\cB\times\cC) \rightarrow (\cB\times\cC)$ be the set of possible modifications to the active architecture at the current step. A choice $c_{i,j}\in\cO$ describes a set update $(\cA_t, S_t)+\{c_{i,j}\}$ where $i\in\{+,-\}$ denotes selection and rejection, and $j\in\{\cA,S\}$ denotes the active actuator and sensor sets, respectively. Unlike the earlier algorithms with finite choices and constraints that limit iterations, we may impose an upper bound $N$ on the maximum number of changes per iteration for each active set. If $N \rightarrow \infty$, then the number of changes is unrestricted. Let $S_1 \> \backslash \> S_2 = S_1 \cap S_2^c$ be the unique elements of $S_1$ not in $S_2$. Then the number of changes from $S_1$ to $S_2$ is given by $\cH(S_1, S_2) = \max\left(\card{S_1 \>\backslash\> S_2},\> \card{S_2 \>\backslash \> S_1}\right)$.


As larger active sets reduce the estimated control cost \eqref{eq:estimated_control_cost}, we may consider selecting $\cL_M$ actuators and $\cL_{M'}$ sensors. However, this results in higher running and switching costs when considering the total predicted cost \eqref{eq:estimated_totalcost}. The greedy optimal architecture might also require less than $N$ changes from the previous time step. To address this, we introduce a possible choice $c_0$ that denotes no update (i.e. $(\cA_t, S_t) + \{c_0\} = (\cA_t, S_t)$) as an option only if the hard constraints on both active sets are satisfied. If all changes in the selection or rejection subsequence increase the total cost, then the greedy optimal sets are the current sets ($c^* = c_0$). If there is no net change to the architecture after sequential selection and rejection in an optimization iteration, then we exit the algorithm (line \ref{alg_step:greedy_sequential_optimal}).



We define the pair of constraints
\begin{align}
    \cL^\text{sel}_{N'}(\cA_t, S_t)&=
    \begin{cases}
        \true & \text{ if } \cL_{m+N'}\leq \card{\cA} \leq \cL_{M+N'} \\ & \text{ and } \cL_{m'+N'}\leq \card{S} \leq \cL_{M'+N'}\\
        \false & \text{otherwise}.
    \end{cases} \nonumber
    \\
    \cL^\text{rej}_{N'}(\cA_t, S_t)&=\cL(\cA_t, S_t) \label{eq:force_constraints}
\end{align}
to force sequential greedy selection and rejection of $N'$ pairs of actuators and sensors by running $2N'$ iterations each of the selection and rejection subsequences. To ensure that the limits of the active sets are preserved at the end of each iteration, we ensure $N'\leq N$, preferably $N'=1$. During each iteration of sequential swapping, there are $N'$ selections and $N'$ rejections for each architecture set during their respective subsequences. To ensure that the modified constraints (to force selection or rejection) are always satisfied at the end of each subsequence, the set of possible modifications $\cO$ is determined by prioritizing architecture sets that fail the hard constraints. If the hard constraints are satisfied, then soft constraints based on the trade-off between control costs and running costs biased by switching costs determine the optimization of the active set. Detailed steps explaining this priority system for simultaneous optimization of two sets during the selection and rejection subsequences are provided in Appendix \ref{ap:choice_algorithms}.

\begin{algorithm}[!ht]
\caption{Architecture optimization with Greedy Swapping}
\label{alg:greedy_sequential_architecture}
\begin{algorithmic}[1]
\Require Available Architecture $\cB, \cC$,
Augmented state estimate $\hat{x}(t)$,
Cost metric $\bJ^{\text{est}}_{t}$ \eqref{eq:estimated_totalcost},
$\cL^\text{sel}_{N'}(\cA_t, S_t)$, $\cL^\text{rej}_{N'}(\cA_t, S_t)$ \eqref{eq:force_constraints},
$(\cA_t, S_t)$ as random selection of $\cB, \cC$ s.t. $\cL(\cA_t, S_t)=\true$ at $t=0$ or previous architecture $(S_{t-1}, S_{t-1})$,
Max number of architecture changes to terminate algorithm $N$,
Max number of selections/rejections per iteration $N'$
\Ensure {Greedy optimal architecture: $(\cA_t, S_t)$}
\State $(\cA^{\text{ref}}_{0}, S^{\text{ref}}_{0}) \gets (\cA_{t}, S_{t})$
\State $n_\text{count}\gets0$
\While{$n_\text{count} < 2N$} \label{alg_step:greedy_sequential_finite_changes}
    \State $n_\text{sel}\gets0$, $(\cA^{\text{ref}}_{1}, S^{\text{ref}}_{1}) \gets  (\cA_t, S_t)$ \label{alg_step:greedy_sequential_selection_start}
    \While{$n_\text{sel} < 2N'$} 
        \State $\cO \gets$ selection choices for $\cL^\text{ref}_{N'}(\cA_t, S_t)$ (Alg. \ref{alg:greedy_selection_choices})
        \State $c^* \gets \argmin_{c\in\cO} \bJ^{\text{est}}_{t}(\hat{x}(t), (\cA_t, S_t) + \{c\})$
        \State $(\cA_t, S_t) \gets (\cA_t, S_t) + \{c^*\}$
        \If{$c^* = c_0$}
            \State \BREAK \Comment{No valuable selections}
        \EndIf
        \State $n_\text{sel} \gets \cH(\cA^{\text{ref}}_{1}, \cA_t) + \cH(S^{\text{ref}}_{1}, S_t)$
    \EndWhile \label{alg_step:greedy_sequential_selection_end}
    \State $n_\text{rej}\gets0$, $(\cA^{\text{ref}}_{2}, S^{\text{ref}}_{2}) \gets (\cA_t, S_t)$ \label{alg_step:greedy_sequential_rejection_start}
    \While{$n_\text{rej} < 2N'$}
        \State $\cO \gets$ rejection choices for $\cL^\text{rej}_{N'}(\cA_t, S_t)$ (Alg. \ref{alg:greedy_rejection_choices})
        \State $c^* \gets \argmin_{c\in\cO} \bJ^{\text{est}}_{t}(\hat{x}(t), (\cA_t, S_t) - \{c\})$
        \If{$c^* = c_0$}
            \State \BREAK \Comment{No valuable rejections}
        \EndIf
        \State $(\cA_t, S_t) \gets (\cA_t, S_t) - \{c^*\}$
        \State $n_\text{rej} \gets \cH(\cA^{\text{ref}}_{2}, \cA_t) + \cH(S^{\text{ref}}_{2}, S_t)$
    \EndWhile \label{alg_step:greedy_sequential_rejection_end}
    \If{$\cH(\cA^{\text{ref}}_{1}, \cA_t) + \cH(S^{\text{ref}}_{1}, S_t) = 0$} \label{alg_step:greedy_sequential_optimal}
        \State \BREAK \Comment{No valuable architecture change}
    \EndIf
    \State $n_\text{count} \gets \cH(\cA^{\text{ref}}_{0}, \cA_t) + \cH(S^{\text{ref}}_{0}, S_t)$
\EndWhile \label{alg_step:greedy_swap_end}
\State \Return Greedy optimal architecture: $(\cA_t, S_t)$
\end{algorithmic}
\end{algorithm}

\subsection{Optimization Parameters}\label{subsec:algorithm_optimization_parameters}

In this subsection, we discuss the key tuning parameters of the proposed greedy optimization algorithm.

\subsubsection{Changes per iteration and per selection or rejection step}
The values of $N$ and $N'$ are directly related to the number of iterations of the algorithm and the number of metric computations and comparisons required per step. Increasing $N$ is expected to increase computation time while yielding diminishing returns, typical of greedy algorithms. If $N\rightarrow\infty$, then an extra iteration loop is required to verify optimality of the active sets. Adjusting $N'$ affects the constraints on the active set and hence the number of iterations within the sequential selection and rejection steps. Increasing $N'$ increases the number of computations per iteration of the sequential algorithm, but can potentially reduce the number of iterations required.

\subsubsection{Prediction horizon}
The length of the prediction horizon $\cT$ affects the frequency of changes in the active sets. A short prediction horizon encourages more aggressive switching of active sets to stabilize the current-state estimate. However, the algorithm can become overly sensitive to disturbances and incur high costs from frequent switching. A longer prediction horizon reduces the switching of active sets and can converge to an architecture within the constraints of relatively low average costs.
However, this decreased sensitivity to disturbances may result in higher control costs, and the self-tuning architecture may not guarantee mean-square stability.



\subsubsection{Architecture constraints and costs}
The available architecture and constraints on active sets define the performance limits of optimal control and estimation. For a cardinality-constrained architecture problem with no running or switching costs, the lower bounds are trivial, and the active sets saturate the network within the limits of their upper bounds. If the ratio of these upper bounds to the number of unstable modes is very small, this can lead to unbounded costs for infinite-horizon stability and control. If the ratio is large, then the problem approaches the trivial case of full controllability and observability.
Imposing switching \eqref{eq:arch_switch_cost} and running \eqref{eq:arch_run_cost} costs adjusts the frequency of change and the size of the active sets. 

\section{Numerical Experiments} \label{sec:empirical_analysis}

In this section, we illustrate the significant potential benefits of the proposed self-tuning network control architecture framework. We first demonstrate greedy self-tuning actuator optimization for an LQR example. We then demonstrate simultaneous self-tuning sensor and actuator optimization using greedy swapping in random networks. The numerical experiments are conducted with and without architecture costs and constraints. The code and problem data for greedy self-tuning LQR actuators can be found in \cite{summers2023SupplementarySoftwareSelfTuning}. The code for self-tuning LQG architectures using greedy swapping and analysis of the optimization parameters can be found in \cite{ganapathy2023ArchitectureSelectionSelfTuning}.






\subsection{Self-tuning LQR: Actuator Selection}

We begin with a self-tuning LQR example. A simple greedy actuator selection heuristic for joint architecture and control design enables a self-tuning architecture to provide dramatically improved performance over a fixed architecture. This improvement is realized even with known linear time-invariant dynamics, with the actuator set selected at each time based on the current state.

\begin{figure}[ht!]
  \centering
  \includegraphics[width=0.85\columnwidth]{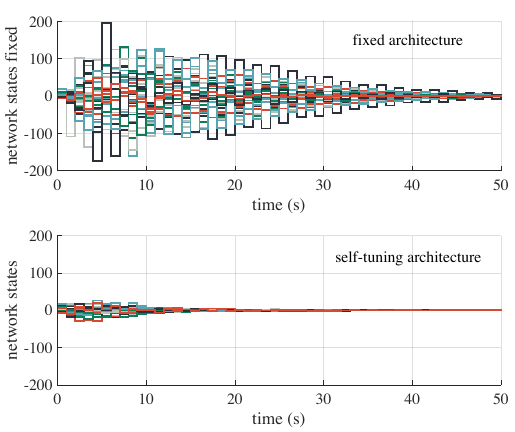}
  \caption{State trajectories for fixed vs self-tuning architecture}
  \label{fig:actuator_tuning_trajectories}
\end{figure}

We consider a 50-node network with a scalar state for each node and a randomly generated unstable dynamics matrix. The set of possible actuators consists of the standard unit basis vectors $\mathcal{B}=\{\mathbf{e}_i \in \bbR^{50}\}$, so each actuator can inject an input signal at each node. The cost matrices are $Q_t = \mbfI_n$ and $R_t = \mbfI_{\card{\cA_t}}$, i.e. the input cost matrix is identical for every set of actuators. The disturbance is i.i.d with $w(t) \sim \mathcal{N}(0, 1e^{-4})$. We randomly generate initial states $x_0\sim\mathcal{N}(0,25)$. The number of actuators available at each time is limited to $K=2$. We simulate network dynamics with a fixed architecture $B=\mat{\mathbf{e}_1 & \mathbf{e_2}}$ using the optimal LQR policy, and with a self-tuning architecture using a greedy heuristic as described in Alg. \ref{alg:greedy_selection_architecture}. The state trajectories are shown in Fig. \ref{fig:actuator_tuning_trajectories}. The cost of the fixed architecture is far worse, a factor 80x more, than the cost of the self-tuning architecture.

\subsection{Self-Tuning LQG: Combined Sensor \& Actuator Selection}

We now evaluate our approach for simultaneous actuator and sensor selection for optimal output feedback control (Section \ref{sec:architecture_optimization_problem_dp_partialfeedback}). We compare fixed vs. self-tuning architectures using the proposed greedy swapping approach (Alg. \ref{alg:greedy_sequential_architecture}). 

We define an $n=50$-node network with a fixed dynamics $A=V\Lambda V^\top$ matrix where the eigenvectors (columns of $V$) are randomly generated orthonormal vectors of $\mathbf{R}^{n}$ and $\Lambda=\diag{\lambda_i(A)}$ is uniformly sampled such that $\lVert |\lambda_i(A)| -1 \rVert \leq 0.1 \> \forall i \in [1, \dots, n]$. The available architectures are $\mathbf{e}_i \in  \bbR^{n}$ with no running or switching costs and tight constraints $\cL_m = \cL_M = \cL_{m'} = \cL_{M'} = 5$. The process and measurement noises are i.i.d with $w(t) \sim \cN(\mbfzero_{n}, \mbfI_{n})$ and $v(t) \sim \cN(\mbfzero_{5}, \mbfI_{5})$. The self-tuning architectures are optimized over a prediction horizon of $T_p=10$ and allow for an unconstrained number of iterations $N\rightarrow \infty$ with one change per architecture set during each selection or rejection subsequence $N'=1$. 
For fixed architectures, only optimal control input and incurred costs are evaluated at each time step. For self-tuning architectures, the greedy optimal architecture, their corresponding optimal gains, and the incurred costs are evaluated at each step.

\begin{figure}[ht!]
  \centering
  \includegraphics[width=0.95\columnwidth]{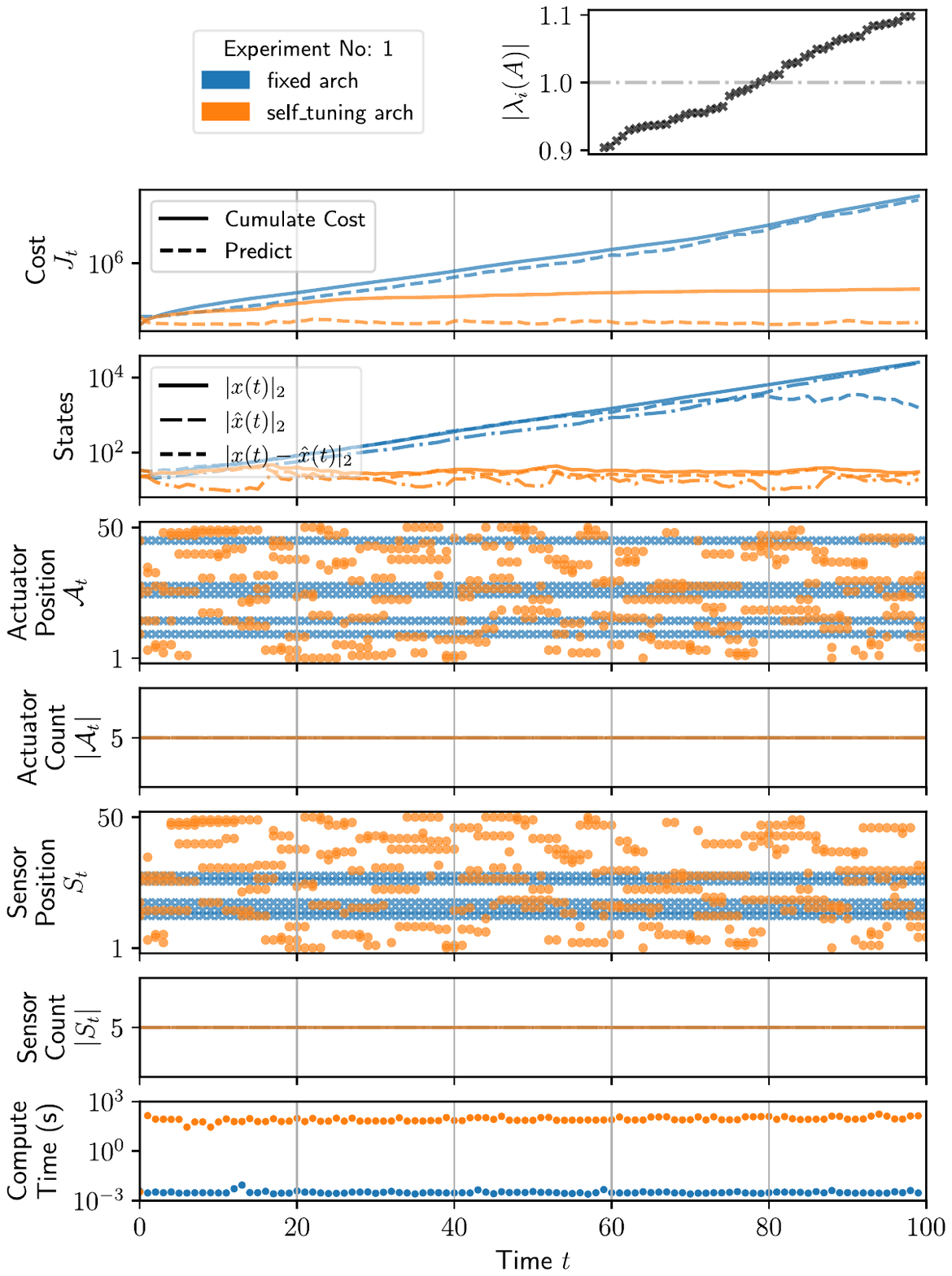}
  \caption{Performance of greedy self-tuning architecture vs fixed architecture. True cumulative costs and predicted costs are lower, norms of the states, state estimates and errors are smaller.}
  \label{fig:single_network_test}
\end{figure}

The results are shown in Fig. \ref{fig:single_network_test}. The plot on the top right shows the distribution of the magnitude of the open-loop eigenvalues of the network dynamics. From top to bottom, the plots show the predicted and cumulative true cost, the norms of the true states, state estimates and error vectors, the location and size of the active actuators and sensors, and computation time per simulation time step. In all plots, the blue and orange elements represent the fixed and self-tuning architecture models, respectively.  Notably, the fixed architecture fails to stabilize the system over the simulation time horizon, whereas the self-tuning architecture attains stability with much lower overall costs. 
The self-tuning architecture achieves better performance by adapting locations of the active sets while having the same set cardinalities as the fixed architecture.




\subsection{Operating Costs and Loose Constraints on Architecture}
\begin{figure}[ht!]
  \centering
  \includegraphics[width=0.95\columnwidth]{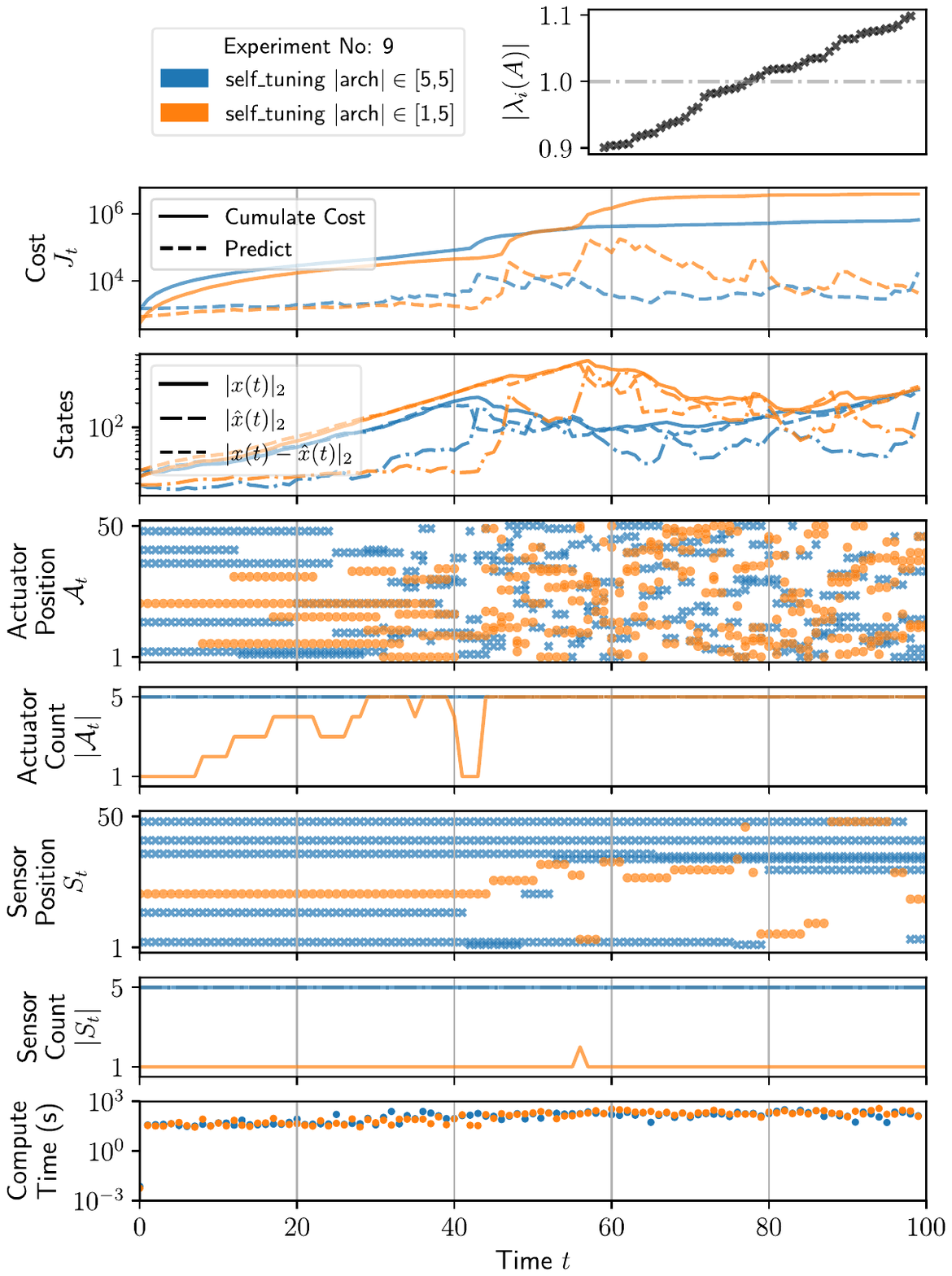}
  \caption{Performance of greedy self-tuning architecture with variable active set size, running and switching costs and limited number of iterations and changes. Changes to the active set size depend on the cost minimization trade-off between predicted control cost and the running and switching costs of the architecture.}
  \label{fig:statistics_selftuning_arch_costs_single}
\end{figure}

We ran additional experiments with running and switching costs ($R_{2,\cB} = R_{3,\cB} = 100\times \mbfone_{|\cB|}, R_{2,\cC} = R_{3,\cC} = 100 \times \mbfone_{|\cC|}$). We limit the
number of iterations to $N=2$ while allowing one change per architecture set $N'= 1$ during each selection or rejection subsequence. We compare the performance of self-tuning architectures under loose set constraints ($\cL_{m}=\cL_{m'}=1 < \cL_{M} = \cL_{M'} = 5$) with the earlier tight constraints. The results are shown in Fig. \ref{fig:statistics_selftuning_arch_costs_single}.

The changing size of the active architecture and the corresponding costs show the effects of architecture costs and loose set constraints. While smaller architecture sets generally have lower predicted costs, they take longer to respond to unexpected disturbances. The size of the active sets decreases if the decrease in $\cJ^{\text{run}}_{t}$ \eqref{eq:arch_run_cost} is more than the increase in $\cJ^{\text{est}}_{t}$ \eqref{eq:estimated_control_cost} and $\cJ^{\text{switch}}_{t}$ \eqref{eq:arch_switch_cost}, as this minimizes the total predicted cost. Conversely active sets increase in size when predicted control costs outweigh running and switching costs.


\section{Conclusion} \label{sec:conclusion}

We introduced a framework for self-tuning network architecture for full-state feedback. We have demonstrated how dynamic programming offers a solution for joint architecture and feedback policy optimization. This concept is then extended to create a framework for assessing the cost of sensors and actuators in discrete-time Linear Time-Invariant (LTI) systems, accounting for additive process and measurement noise. To tackle the combinatorial optimization problem within this framework, we have presented a computationally efficient variation of the greedy algorithm. Numerical tests have confirmed the cost improvement achieved by our self-tuning architecture optimization algorithm. The ability to fine-tune the algorithm's parameters for specific applications opens up possibilities for future research in control optimization, combining elements of stochastic adaptive control, system identification, reinforcement learning, and architecture design, exploring the trade-off between exploration and exploitation.

\pagestyle{empty}
\bibliographystyle{IEEEtran}
\bibliography{SelfTuningArchitecture}

\begin{IEEEbiography}[{\includegraphics[width=\textwidth]{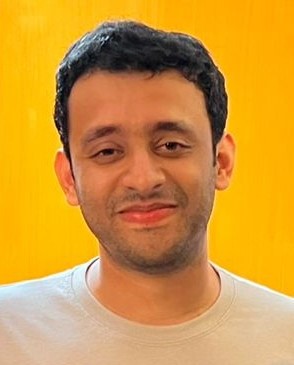}}]
{Karthik Ganapathy} received the B.Tech. degree in mechanical engineering from Amrita School of Engineering, Coimbatore, India, in 2016. He is currently a Ph.D candidate at The University of Texas at Dallas, Richardson, USA. His research interests include architecture performance bounds, model-based optimization, network control and estimation, and online heuristic algorithms.
\end{IEEEbiography}
\begin{IEEEbiography}[{\includegraphics[width=\textwidth]{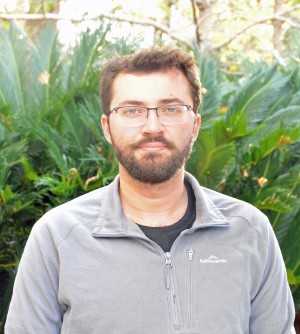}}]
{Iman Shames} received the B.Sc. degree in electrical engineering from Shiraz University, Shiraz, Iran, in 2006, and the Ph.D. degree in engineering and computer science from The Australian National University, Canberra, ACT, Australia in 2011.,He was an ACCESS Post-Doctoral Researcher with the ACCESS Linnaeus Centre, KTH Royal Institute of Technology, Stockholm, Sweden. He was an Associate Professor with the Department of Electrical and Electronic Engineering, University of Melbourne, Melbourne, VIC, Australia, from 2014 to 2020, where he was a Senior Lecturer and a McKenzie Fellow from 2012 to 2014. He is currently a Professor of Mechatronics with the School of Engineering, The Australian National University. His current research interests include, but are not limited to, optimization theory and its application in control and estimation, mathematical systems theory, and security and privacy in cyber–physical systems.
\end{IEEEbiography}
\begin{IEEEbiography}[{\includegraphics[width=\textwidth]{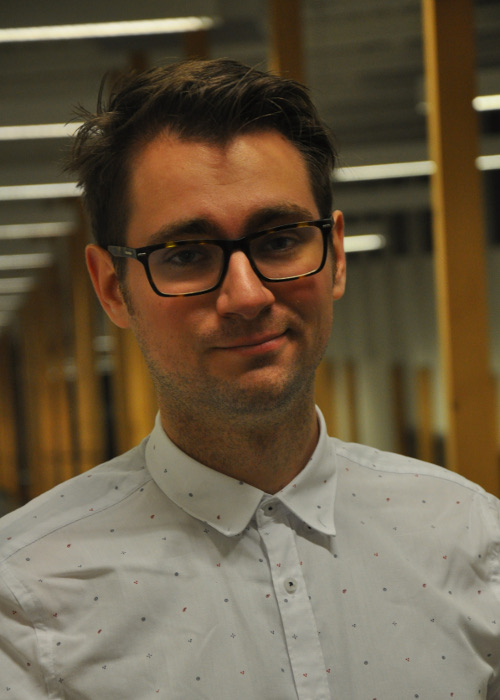}}]
{Mathias Hudoba de Badyn}
is an Associate Professor in the Department of Technology Systems at the University of Oslo. From 2019-2023, he was a postdoctoral scholar in the Automatic Control Laboratory (IfA) at the Swiss Federal Institute of Technology (ETH), Z\"{u}rich. He received both his Ph.D.~degree in the William E.~Boeing Department of Aeronautics and Astronautics, and an M.Sc.~degree in the Department of Mathematics in 2019 at the University of Washington. In 2014, he graduated from the University of British Columbia with a BSc in Combined Honours in Physics and Mathematics. His research interests include the analysis and control of networked dynamical systems, with applications in aerospace vehicles and energy infrastructure.
\end{IEEEbiography}

\begin{IEEEbiography}[{\includegraphics[width=\textwidth]{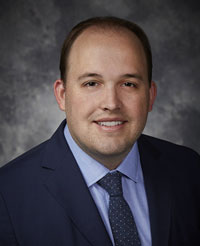}}]
{Tyler Holt Summers} received the B.S. degree in mechanical engineering from Texas Christian University, Fort Worth, TX, USA, in 2004, and the M.S. and Ph.D. degrees in aerospace engineering with emphasis on feedback control theory from the University of Texas at Austin, Austin, TX, USA, in 2007 and 2010, respectively. He is currently an Associate Professor of mechanical engineering with an affiliate appointment in electrical engineering with the University of Texas at Dallas, Richardson, TX. Prior to joining UT Dallas, he was an ETH Postdoctoral Fellow with the Automatic Control Laboratory, ETH Zürich, from 2011 to 2015. He was a Fulbright Postgraduate Scholar with the Australian National University, Canberra, Australia, in 2007–2008. His research interests include feedback control and optimization in complex dynamical networks, with applications to electric power networks and distributed robotics.
\end{IEEEbiography}


\newpage

\appendices

\section{Proof for Theorem 1} \label{proof:thm1}

The proof follows the same inductive argument as in standard dynamic programming \cite{smith1996DynamicProgrammingOptimal}, except that the minimization step is performed jointly over architectures and policies. For each time $t \in [0, \cT-1]$, we define the tail architecture policy $\pi^t = [\pi_t, \pi_{t+1}, \dots, \pi_{\cT+1}]$, where $\pi_t$ specifies $(\cA_t, \overline{\pi}_t)$ with $u^{\cA_t}(t)= \overline{\pi}(x(t))$. We then define the optimal cost-to-go functions at time $t$ for state $x(t)=x$
\begin{align*}
    J_t^*(x) = \min_{\pi_t} \bbE \left[ \sum_{\tau=t}^{\cT-1} c_\tau (x(\tau), \overline{\pi}_t(x(\tau)) + c_{\cT}(x(\cT))  \right]
\end{align*}
with expectation with respect to tail disturbance sequences. We define $J^*_\cT(x) = c_\cT(x)$, establishing the inductive base case. Now assume for some $t$ that $J^*_{t+1}(x) = J_{t+1}(x)$. We have
\begin{align*}
    J_t^*(x) &= \min_{\pi_t} \bbE_{w} [ c_t(x, \overline{\pi}_t(x)) + \nonumber\\&\quad\quad\quad [\min_{\pi^{t+1}} \bbE[\sum_{\tau=t+1}^{\cT-1} c_\tau(x(\tau) ,\overline{\pi}_\tau(x(\tau))) ]]] \nonumber\\
    &= \min_{\pi_t} [\bbE_{w} c_t(x, \overline{\pi}_t(x)) + J^*_{t+1}(x^+)] \nonumber\\
    &= \min_{\pi_t} [\bbE_{w} c_t(x, \overline{\pi}_t(x)) + J^*_{t+1}(f_{\theta_t, \cA_t}(x, \overline{\pi}_t(x), w))] \nonumber\\
    &= \min_{\pi_t} [\bbE_{w} c_t(x, \overline{\pi}_t(x)) + J_{t+1}(f_{\theta_t, \cA_t}(x, \overline{\pi}_t(x), w))] \nonumber\\
    &= \min_{\genfrac {}{}{0pt}{}{(\cA_t, u)\in \cB_\cK \times}{\prod_{u_i \in \cA_t}\cU_i}} \bE_w \left[ c_t(x,u^{\cA_t}) + J_{t+1}(f_{\theta_t, \cA_t}(x, u^{\cA_t}, w)) \right] \nonumber\\
    &= J_t(x) \nonumber
\end{align*}
where the first step follows from the Principle of Optimality and Markovian structure of the dynamics, the second by definition of $J^*_{t+1}$, the third by the state update equation, the fourth by the induction hypothesis, and the fifth by definition of the architecture-policy. This establishes the induction step and completes the proof.

\section{Proof for Theorem 2} \label{proof:thm2}

The dynamic programming recursion for \eqref{eq:thm2_problem} is
\begin{align*}
    J_\cT &= x^\top Q_\cT x\\
    J_t(x) &= \min_{(u, B^{\cA_t}) \bbR^K\times \bU_K} \bbE[x^\top Q_t x + u^\top R_t u + J_{t+1}(x^+)],
\end{align*}
where $x^+ = Ax + B^{\cA_t}u + w$. We simplify the notation for the architecture optimization variable $B_t:=B^{\cA_t}$. The recursion one time step backwards from $\cT$ can be split into a minimization over $u$ and over $B_{\cT-1}$ as,
\begin{align*}
    J_{\cT-1}(x) = &\min_{B_{\cT-1}\in\bU_K} \min_{u\in\bbR^K} \bbE[x^\top Q_{\cT-1}x + u^\top R_{\cT-1}u \\& + (Ax+B_{\cT-1}u+w)^\top Q_\cT (Ax+B_{\cT-1}u+w)].
\end{align*}

Evidently, the inner minimization over $u$ yields
\begin{align*}
    u^*_{\cT-1}(x) = -(B_{\cT-1}^\top Q_\cT B_{\cT-1} + R)^{-1} B_{\cT-1}^\top Q_\cT Ax.
\end{align*}

Substitution gives us a quadratic function to minimize. For a finite number of choices of $B_{\cT-1}$, each depending on $x$, it follows that $J_{\cT-1}(x)$ is piecewise quadratic and $u^*_{\cT-1}(x)$ piecewise linear.

The remainder of the proof can be done by induction on a standard ansatz, as follows. Suppose $J_t(x) = x^\top P_t(B_t(x))x + q_t(B_t(x))$, where $P_t(B_t(x))=:P_t^{B_t}$ and $q_t(B_t(x))$ depend on the choice of $B_t$ at time $t$, which in turn depends on $x$; in other words, $J_t(x)$ is piecewise quadratic. We can write
\begin{align*}
    J_{t-1}(x) &= \min_{B_{t-1}\in\bU_K} \min_{u\in\bbR^K} \left[ \trace{P_t^{B_t}W} + q_t(B_t(x^+)) \right. \\  + \mat{ x \\ u }^\top &
    \mat{
        A^\top P_t^{B_t}A + Q_{t-1} & A^\top P_t^{B_t} B_{t-1} \\
        B_{t-1}^\top P_t^{B_t} A & B_{t-1}^\top P_t^{B_t} B_{t-1} + R_{t-1}
    } \left. \mat{ x \\ u } \right]\\
    = &\min_{B_{t-1}\in\bU_K} q_t(B_t(x^+)) + \trace{P_t^{B_t}W} + x^\top \left[A^\top P_t^{B_t}A + \right. \\Q_{t-1} - A^\top&P_t^{B_t} B_{t-1}(B_{t-1}^\top P_t^{B_t} B_{t-1} + R_{t-1})^{-1}B_{t-1}^\top P_t^{B_t} A \left. \right] x \\
    = \quad& \min_{B_{t-1}\in\bU_K} \left[ x^\top P_{t-1}^{B_{t-1}} x \right] + q_{t-1}(B_{t-1}(x)),
\end{align*}
where $\bbE[w(t)^\top P_{t+1}(B_t(x))w(t)] := \trace{P_t^{B_t} W}$, $\trace{P_t^{B_t} W} + q_t(B_t(x^+)) := q_{t-1}(B_{t-1}(x)),$ and where the problem can be solved by exhaustive search over the $\genfrac{(}{)}{0pt}{}{n}{k}$ elements of $\bU_K$. It follows that the function is piecewise continuous in $x$ and piecewise quadratic.

\section{Priority Choices for Selection and Rejection Subsequences} \label{ap:choice_algorithms}

The greedy approximation of combinatorial swapping is achieved by sequential selection and rejection subsequences. Swapping is forced by adjusting the hard constraints on the active sets.. When optimizing two sets simultaneously using a single metric, we require a priority system to determine the set of choices $\cO$ to ensure that the hard constraints are satisfied after all iterations of the selection or rejection subsequence. We describe the priority system and the corresponding algorithm for each subsequence below.

\subsection{Priority selection as described in Alg. \ref{alg:greedy_selection_choices}}
\begin{itemize}
\item High-priority selection choices:
    \begin{itemize}
        \item If $\cL_{m+N'} > \card{\cA_t}$, then actuators \textbf{must} be considered for selection
        \item If $\cL_{m'+N'} > \card{S_t}$, then sensors \textbf{must} be considered for selection
    \end{itemize}
\item Consider low-priority selection choices \textbf{only if} no high priority selection choices
    \begin{itemize}
        \item If $\card{\cA_t} < \cL_{M+N'}$, then actuators may be considered for selection
        \item If $\card{S_t} < \cL_{M'+N'}$, then sensors may be considered for selection
        \item If $\cL^\text{sel}_{N'}(\cA_t, S_t)$ is satisfied, then no update $c_0$ may be considered
    \end{itemize}
\end{itemize}
\begin{algorithm}[ht!]
\caption{Choices for Forced Selection based on Priority}
\label{alg:greedy_selection_choices}
\begin{algorithmic}[1]
\Require Current active architecture $\cA_t, S_t$, Available architecture $\cB, \cC$, Selection constraints $\cL^\text{sel}_{N'}(\cA_t, S_t)$
\Ensure Selection choices $\cO$
\State $\cO \gets \emptyset$
\If{$\cL_{m+N'} > \card{\cA_t}$ or $\cL_{m'+N'} > \card{S_t}$}
    \If{$\cL_{m+N'} > \card{\cA_t}$} 
        \State $\cO \gets \cO + \{c_{+,S} \mid c \in \cA_t^c\}$
    \EndIf
    \If{$\cL_{m'+N'} > \card{S_t}$} 
        \State $\cO \gets \cO + \{c_{+,S} \mid c \in \cA_t^{'c}\}$
    \EndIf
\Else 
    \If{$\card{\cA_t} < \cL_{M+N'}$}
        \State $\cO \gets \cO + \{c_{+,S} \mid c \in \cA_t^c\}$
    \EndIf
    \If{$\card{S_t} < \cL_{M'+N'}$}
        \State $\cO \gets \cO + \{c_{+,S} \mid c \in \cA_t^{'c}\}$
    \EndIf
    \If{$\cL^\text{sel}_{N'}(\cA_t, S_t)=\true$}
        \State $\cO \gets \cO + \{c_0\}$
    \EndIf
\EndIf
\State \Return Selection choices $\cO$
\end{algorithmic}
\end{algorithm}

\subsection{Priority rejection as described in Alg. \ref{alg:greedy_rejection_choices}}
\begin{itemize}
\item High-priority rejection choices:
    \begin{itemize}
        \item If $\cL_{M} < \card{\cA_t}$, then actuators \textbf{must} be considered for rejection
        \item If $\cL_{M'} < \card{S_t}$, then sensors \textbf{must} be considered for rejection
    \end{itemize}
\item Consider low-priority rejection choices \textbf{only if} no high priority rejection choices
    \begin{itemize}
        \item If $\card{\cA_t} > \cL_{m}$, then actuators may be considered for rejection
        \item If $\card{S_t} > \cL_{m'}$, then sensors may be considered for rejection
        \item If $\cL^\text{rej}_{N'}(\cA_t, S_t)$ is satisfied, then no update $c_0$ may be considered
    \end{itemize}
\end{itemize}

\begin{algorithm}[ht!]
\caption{Choices for Forced Rejection}
\label{alg:greedy_rejection_choices}
\begin{algorithmic}[1]
\Require Current active architecture $\cA_t, S_t$, Available architecture $\cB, \cC$, Selection constraints $\cL^\text{rej}_{N'}(\cA_t, S_t)$
\Ensure Rejection choices $\cO$
\State $\cO \gets \emptyset$
\If{$\cL_{M} < \card{\cA_t}$ or $\cL_{M'} < \card{S_t}$}
    \If{$\cL_{M} < \card{\cA_t}$}
        \State $\cO \gets \cO + \{c_{-,S} \mid c \in \cA_t\}$
    \EndIf
    \If{$\cL_{M'} < \card{S_t}$}
        \State $\cO \gets \cO + \{c_{-,S} \mid c \in S_t\}$
    \EndIf
\Else 
    \If{$\card{\cA_t} > \cL_{m}$}
        \State $\cO \gets \cO + \{c_{-,S} \mid c \in \cA_t\}$
    \EndIf
    \If{$\card{S_t} > \cL_{m'}$}
        \State $\cO \gets \cO + \{c_{-,S} \mid c \in S_t\}$
    \EndIf
    \If{$\cL^\text{rej}_{N'}(\cA_t, S_t)=\true$}
        \State $\cO \gets \cO + \{c_0\}$
    \EndIf
\EndIf
\State \Return Rejection choices $\cO$
\end{algorithmic}
\end{algorithm}

\end{document}